\documentclass[runningheads]{llncs}

\usepackage[T1]{fontenc}
\usepackage{graphicx}
\usepackage{booktabs}
\usepackage{amsmath}
\usepackage{amssymb}
\usepackage{array}
\usepackage{tabularx}
\usepackage{microtype}   
\usepackage{marvosym}
\usepackage{url}         
\begin{document}

\title{Engagement-Led Segmentation of Gamified Participation Data in a Large-Scale Remote Internship: A Mixed-Methods Study}
\titlerunning{Engagement-Led Segmentation of Gamified Participation Data}

\author{Sakshi Sharma\inst{1}(\Letter) \and
Pavani Ayinampudi\inst{2} \and
Aditya B.M.V.\inst{2} \and
Jinal Gupta\inst{2} \and
Prakash Hegade\inst{2} \and
Rohit Sharma\inst{1} \and
Meenakshi V\inst{1} \and
S.R.S. Iyengar\inst{1}}
\authorrunning{S. Sharma et al.}
\institute{Indian Institute of Technology Ropar, Rupnagar, Punjab, India\\
\email{\{sakshi.23csz0006,rohit.24csz0014\}@iitrpr.ac.in}\\
\email{\{meenakshi.19csz0013,sudarshan\}@iitrpr.ac.in}
\and
ANNAM.AI, Indian Institute of Technology Ropar, Rupnagar, Punjab, India\\
\email{\{pavania.harvard2025,adityabmv,jinalbirla,prakash.hegade\}@gmail.com}}

\maketitle

\begin{abstract}
Gamified points are widely used to represent learner participation, but cumulative point totals provide limited information about how participation changes over time. This limitation matters in continuously enrolling programmes, where learners have different opportunities to accumulate points. This study examines how gamified participation data can be interpreted as indicators of behavioural engagement in a large-scale, continuously enrolling remote internship. Using an anonymised operational dataset, 3{,}607 distinct started learners were first classified by participation status, separating dormant learners from those with observable participation. The 1{,}871 active learners were then grouped using $K$-means clustering on normalised longitudinal and opportunity-aware participation features. Four participation patterns emerged: Thriving Core, Tapering, Fast Starters, and Occasional Participants. Their trajectories differed in both level and direction. A perception survey of 597 respondents provided complementary learner-reported evidence, integrated with the behavioural strand through a joint display by segment. Perceptions differed across segments, while individual-level correlations between perceptions and behaviour were small, and learners with low recorded participation could report positive perceptions of the points system alongside external barriers. Segments assigned from the first four weeks were then checked against eleven weeks of later platform records: 94\% of the Thriving Core attended at least ten further sessions and 22\% completed the internship, against under 7\% and under 1\% in the low-engagement segments. The findings suggest that gamified participation data are more informative when interpreted as longitudinal, opportunity-aware behavioural indicators rather than cumulative scores alone.

\end{abstract}

\section{Introduction}\label{sec:intro}

Engagement is a fundamental part of learning, shaping how learners interact with, respond to, and invest themselves in the learning process. Learner engagement broadly refers to the level of involvement learners show in their learning and is commonly understood through behavioural, emotional, and cognitive dimensions. Among these dimensions, behavioural engagement provides visible indications of learners' involvement through actions such as attending sessions, responding to questions, completing tasks, and participating in discussions. Such behaviours become particularly important in large-scale learning programmes, where direct observation of individual learners is difficult. Digital learning environments can capture these behaviours as participation records, creating opportunities to make learner engagement visible and measurable.

Digital learning environments commonly use gamified points tables to represent and encourage learner participation~\cite{deterding2011,koivisto2019}. Points are typically awarded for observable activities such as attending sessions, responding to polls, or completing tasks. These activities provide behavioural traces of participation, but they do not capture all dimensions of learner engagement, which also includes cognitive and emotional involvement that cannot be directly observed~\cite{fredricks2004}. Therefore, gamification points should be interpreted as a behavioural proxy for learner engagement rather than as a direct measure of engagement~\cite{sailer2020}.

Point totals provide a simple summary of learner participation, but they do not show how participation changes over time. Two learners may earn the same number of points but follow very different participation patterns~\cite{kizilcec2013}. One learner may participate consistently throughout the programme, another may become inactive after an active start, while a third may gradually increase participation. These differences are not visible in a cumulative total. Despite the widespread use of gamification, how cumulative point totals should be interpreted as indicators of learner engagement remains less explored.

This challenge becomes particularly important in remote internship programmes, where learner participation is recorded mainly through digital platforms. In face-to-face settings, mentors can directly observe learners' participation and interaction, whereas remote programmes rely more heavily on behavioural traces such as attendance records, meeting participation, and poll responses~\cite{henrie2015,siemens2011}. The growing emphasis on internship-based learning through initiatives such as the National Education Policy 2020~\cite{nep2020}, the AICTE Internship Policy~\cite{aicte2019}, and the Prime Minister Internship Scheme~\cite{pmis2024} further highlights the need to monitor learner participation at scale. This need is particularly relevant for continuously enrolling programmes, where learners join at different times and therefore have different opportunities to accumulate points.

Examining participation as a trajectory rather than a cumulative total provides an opportunity to understand how learner engagement changes during a programme. Instead of asking how many points a learner has earned, a trajectory-based view asks whether participation is increasing, remaining stable, or declining over time. This study explores behavioural participation trajectories using routinely collected gamification data from a large-scale remote internship programme and reads the resulting trajectories together with a learner perception survey. The segmentation uses ongoing behavioural indicators from the first four weeks; a later export of the same platform is then used to check the segments against subsequent participation and completion. The study aims to improve the interpretation of gamification points for monitoring learner engagement in continuously enrolling internship programmes.

\section{Background Study}

Learner engagement has three dimensions: behavioural, emotional, and cognitive~\cite{fredricks2004}. In remote settings, behavioural engagement is the most directly observable because it leaves traces on the platform, such as attendance records, participation logs, and poll responses~\cite{henrie2015}. In this study, attendance and poll participation are treated as indicators of participation in programme activities, not as direct measures of motivation or learning. Remote internships are a form of work-integrated learning in which regular participation is an important part of the learning experience~\cite{jackson2015}. At the same time, large online programmes can experience substantial early attrition, with many registrants not progressing to sustained participation~\cite{reich2019}. Identifying changes in participation early, before final outcomes are available, is therefore valuable.

To make participation visible and to reward it, many programmes use gamification, defined as the use of game design elements in non-game contexts~\cite{deterding2011}. Points and leaderboards are among the most common elements~\cite{khaldi2023,koivisto2019}. The theoretical grounding of gamification varies, and much of the conventional wisdom about it remains unsettled~\cite{dichev2017}. Reviews report that its effects are mixed and depend on context: they are positive but conditional, often small, and weaker for motivational outcomes than for cognitive ones~\cite{hamari2014,sailer2020}. Design choices matter more than the rewards themselves~\cite{adams2022}. Motivation research adds further caution. Salient extrinsic rewards can undermine intrinsic motivation~\cite{deci1999}, gamification sustained over a semester has been linked to declining motivation~\cite{hanus2015}, and individual game elements can increase output without changing motivation~\cite{mekler2017}. A point total is therefore a joint product of rule design, opportunity, and behaviour. It can reward surface participation when learners optimise for score, and it can appear unfair when opportunities to earn are unequal.

Behavioural trace data provide an opportunity to examine learner participation beyond cumulative point totals. Learning analytics uses such trace data to understand learner participation and identify patterns that may support timely intervention~\cite{siemens2011,blumenstein2020,kizilcec2017}. In large-scale online learning, researchers have shown that learners follow distinct engagement trajectories, and that analysing subpopulations rather than averages can provide a more informative view of learner behaviour~\cite{kizilcec2013,ferguson2015,coffrin2014}. Longitudinal analysis has therefore been recognised as important for understanding how participation changes over time~\cite{knight2017}. However, less attention has been given to interpreting gamification point data as longitudinal indicators of participation, particularly in continuously enrolling programmes where learners have different opportunities to accumulate points. Behavioural traces also cannot explain learners' perceptions or the reasons underlying their participation patterns, highlighting the value of considering learner-reported experiences alongside behavioural trajectories.

\section{Methodology and Methods}\label{sec:methods}

\subsection{Research Questions}
The study is guided by a main research question. \emph{RQ: In a large-scale, continuously enrolling remote internship, how can gamified participation data be interpreted as indicators of behavioural engagement, particularly when learners have different opportunities to participate?} It is addressed through four sub-questions:
\begin{description}
\item[RQ1.] What distinct participation patterns emerge when learners are grouped using normalised, longitudinal trajectories rather than cumulative point totals?
\item[RQ2.] To what extent do learners' perceptions of the points system align with their behaviourally derived segments?
\item[RQ3.] When behavioural patterns and learner perceptions diverge, what reasons and barriers do learners report, particularly among learners with low recorded participation?
\item[RQ4.] Do the segments identified from a learner's first four weeks predict what that learner does over the following eleven weeks (continued attendance, peer-teaching activity, project work, and internship completion), and does the shape of the trajectory predict these outcomes beyond the number of points earned in the same period?
\end{description}
RQ1 to RQ3 use records up to the segmentation snapshot of 29 June 2026. RQ4 is the validation question: it uses only records dated after that snapshot.

\subsection{Research Design and the Two-Tier Model}\label{sec:design}
\begin{figure}[tbp]
\centering
\fbox{\includegraphics[width=0.9\textwidth]{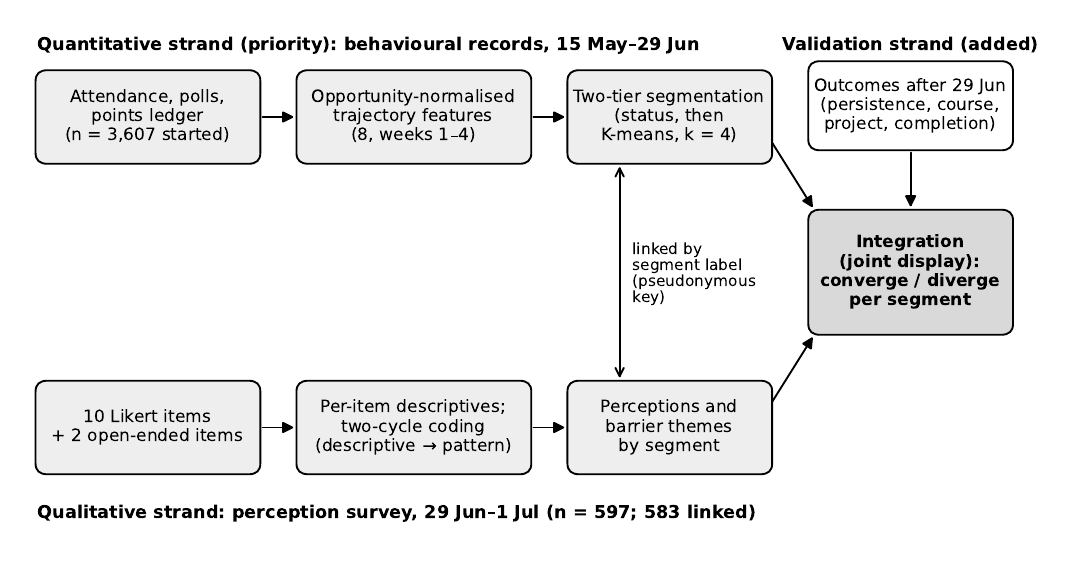}}
\caption{The two strands of the design and their point of integration. The behavioural strand (top) produces the segments; the survey strand (bottom) produces perceptions and barrier themes by segment; the two meet in the joint display through the segment label. The validation strand (top right) reads outcomes recorded after the segmentation snapshot.}\label{fig:integration}
\end{figure}

This study used a convergent mixed-methods design~\cite{creswell2018} in which the quantitative and qualitative strands were collected during the same period, analysed separately, and brought together at the interpretation stage (Fig.~\ref{fig:integration}). The quantitative strand, which was the primary strand, used an observational learning-analytics design based on the platform's own records of attendance, poll responses, and the points ledger to describe what participation looked like across learners. The qualitative strand used a perception survey comprising ten Likert-scale items and two open-ended items to examine learners' perspectives on their participation; the open-ended responses were analysed using two-cycle coding following Salda\~na~\cite{saldana2013}, with descriptive coding in the first cycle against a documented codebook and pattern coding in the second cycle to develop themes~\cite{braun2006}. Two coders independently coded a 20\% subsample, with inter-coder agreement of $\kappa{=}0.88$ for the effect item and $\kappa{=}0.91$ for the barrier item. The convergent design was used because behavioural records indicate what learners did but do not reveal what they thought, whereas survey responses provide learners' perceptions but represent a self-selected group; examining the two strands segment by segment therefore allows areas of agreement and divergence to be identified. Integration was performed through the segment label: each survey response was linked to one learner and, through a separately held key, to one segment. A joint display (Table~\ref{tab:joint}) brings each segment's learner perceptions and barrier themes together with its behavioural profile (Table~\ref{tab:fingerprint}) and later outcomes, and reports a convergence verdict for each segment. The study is not a randomised experiment and therefore makes no causal claims about gamification. The analysis is descriptive and diagnostic. A two-tier model is used to distinguish participation status before examining variation within active learners.

\emph{Tier~1: Participation status.} The registered population is first divided into three groups: not onboarded, dormant, and active. Learners with no recorded start date are classified as not onboarded and excluded from further analysis. Learners who started but had no recorded attendance or poll participation are classified as dormant. Learners with at least one attended session or poll response are classified as active. This separation ensures that learners with no recorded participation are not clustered together with learners who have observable participation trajectories.

\emph{Tier~2: Active-learner segmentation.} $K$-means clustering~\cite{macqueen1967} is applied only to the active learners using standardised longitudinal and participation features. The eight features comprise four weekly velocity measures, one week-1-to-2 acceleration measure, attendance rate, poll participation rate, and points earned per eligible session. Weekly values beyond a learner's observed tenure are treated as zero; 18\% of active learners had fewer than four aligned weeks, so late-week velocity measures may be affected by right-censoring. All features are $z$-scored before clustering.

The number of clusters is selected using both silhouette scores~\cite{rousseeuw1987} and the interpretability of the resulting participation patterns. Silhouette values are 0.59 for $k{=}2$ and 0.47, 0.44, and 0.47 for $k{=}3$, 4, and 5, respectively. Although $k{=}2$ has the highest silhouette, it merges distinct rising and tapering participation patterns. At $k{=}5$, the tapering group is split into two smaller variants. Thus, $k{=}4$ provides the smallest solution that preserves the substantive distinction between rising and tapering participation. The four-cluster solution is stable across alternative initialisations and resampling: the adjusted Rand index (ARI)~\cite{hubert1985} is approximately 1.00 across 100 random seeds and 0.96 across 200 bootstrap samples. Re-clustering in a PCA space explaining 95\% of the variance produces an ARI of 0.98, while clustering on shape-related features alone produces an ARI of 0.85. Agreement with alternative algorithms is moderate (Ward: 0.58; Gaussian mixture: 0.39). The four archetypes therefore provide one useful representation of the active tier rather than fixed learner types. Solutions with two to six clusters were compared using the same features (supplementary Table~Z). The two-cluster solution separates learners with sustained engagement from the remaining learners and predicts later attendance nearly as well as the four-cluster solution (AUC 0.90 vs.\ 0.94). We selected four clusters because it is the smallest solution that distinguishes learners with rising engagement from those whose engagement gradually tapers off, which is the distinction examined in this study. The resulting labels describe observed participation behaviour and weekly shape, not psychological types. Full robustness checks, together with the survey instrument and codebook, are provided in the supplementary technical report.\footnote{Supplementary report: \url{https://drive.google.com/file/d/1OuRtBTLGX5Zp7XLAcKqmT78RBgsZyoxO/view}}

\subsection{Perception Survey and Linkage}
A perception survey consisting of ten 5-point Likert items and two open-ended items was used to capture learners' views of the points system. The Likert items covered usefulness, motivational relevance, self-monitoring, fairness, clarity, consistency, encouragement of attendance and poll participation, and points-related stress. The stress item was negatively valenced and was not reverse-scored; higher scores indicate greater stress. The open-ended items captured learners' experiences and barriers related to participation.

The survey was presented as an optional dashboard prompt in late June 2026 and closed on 1 July 2026. Because the prompt was shown only to learners who visited the dashboard, participation in the survey involved two stages of self-selection: visiting the dashboard and choosing to respond. Dashboard visitation was itself related to learner participation, creating the possibility of differential survey response across the behaviourally derived segments. Each response was linked to the learner's participation segment using a separately held key that mapped email addresses to pseudonyms and then to segments. The email address was discarded from the analytic table immediately after linkage, so the subsequent analysis used pseudonymised records.

Response rates were 83\% in the Thriving Core (411 of 497), 35\% in Tapering (101 of 287), and below 3\% in the remaining three groups: Fast Starters (8 of 303), Occasional (18 of 784), and Dormant (45 of 1{,}736). The survey therefore represents the more engaged learners well but provides limited representation of learners with lower engagement. Accordingly, the differences observed between segments among survey respondents are reported as findings, while comparisons involving the Occasional and Fast Starter groups are treated as exploratory and are not used to draw conclusions. To check whether the results were driven by the small numbers in these groups, all segment comparisons were repeated after pooling the three low-engagement groups ($n{=}71$). The same eight of ten items remained significant after correction, while fairness and barriers remained non-significant, indicating that the observed gradient was not dependent on the two smallest groups.

\section{Data Collection and Analysis}

\subsection{Study Context and the Points System}
This study was conducted at the Vicharanashala Lab for Education Design, Indian Institute of Technology Ropar, within a large-scale, continuously enrolling remote internship. The programme is delivered through synchronous online sessions and poll-based activities, with participation recorded through a gamified points system. Admission involves an automated screening interview, after which learners join the programme on a rolling basis. At the time of the segmentation snapshot the programme was ongoing, so completion outcomes were not available for the segmentation itself; they were read from a later export for RQ4 (Section~\ref{sec:data}). Informed consent was obtained from all participating students for the use of their data for research purposes.

The points system converts participation records into points using predefined rules. During the observation window, the points ledger contained three types of awards: a one-time onboarding grant of 100 points, attendance awards (24{,}800 events; mean $\approx$ 8.5 points per event), and poll participation awards (24{,}156 events; mean $\approx$ 6.3 points per event). No award carried a negative point value; all 48{,}956 recorded events contributed positive points. Points were displayed through a dashboard leaderboard and running balance and had no effect on marks or certification. They therefore served primarily as a form of participation recognition and ranking. The number of opportunities to earn points varied over time: sessions were held on most weekdays, with approximately seven eligible sessions per learner-week during the early period, while the number of polls per session varied from approximately two to ten. Thus, learners did not have a constant number of opportunities to accumulate points.

\subsection{Data Sources and the Participation Funnel}\label{sec:data}
The analysis uses an anonymised operational export generated in late June 2026 from the live programme database (Table~\ref{tab:data}). Because the programme follows a rolling-entry model, the available population represents a participation funnel rather than a single cohort. The applicant roster contained 8{,}479 candidates who had completed the admission interview by the time of the snapshot, of whom 3{,}717 had started the programme. After duplicate accounts were identified, 3{,}607 distinct started learners remained. The student roster contained 7{,}945 accounts corresponding to 7{,}790 distinct persons. Of these, 4{,}183 persons had no recorded start date and had not yet entered the programme. They therefore had no opportunity to generate participation traces and were excluded from all trajectory analyses. The remaining learners formed the population from which participation status and subsequent behavioural patterns were examined.

Learners joined the programme on a rolling basis between 15 May and 29 June 2026 and were expected to complete the same phases. Between 30 June and 5 July, 741 learners were re-registered with new start dates and their points balances reset to 100. As part of this process, their earlier attendance and poll credits were removed from the live points ledger. All analyses in this paper use the 29 June export, which retains the learners' original start dates and complete records from May to June. For RQ4, a second export of the same platform covering 15 May to 14 September 2026 supplied the later records; learners were matched across the two exports through the pseudonym key, so the 741 re-registered learners remain included in the outcome analysis.

\begin{table}[tbp]
\caption{Principal datasets in the anonymised operational export (late-June 2026 snapshot).}\label{tab:data}
\footnotesize
\begin{tabularx}{\textwidth}{@{}llr>{\raggedright\arraybackslash}X@{}}
\toprule
Dataset & Unit & Rows & Use in study \\
\midrule
Student master & Student & 7{,}790 & Status, cohort timing, totals, level/league \\
Normalised metrics & Student & 7{,}790 & Rates, velocity, acceleration, normalised ranks \\
Attendance records & Student-session & $\approx$129{,}000 & Attendance minutes, percentage, qualification \\
Poll participation & Student-session & $\approx$112{,}600 & Poll attempts, missed questions, percentage \\
Point ledger & Transaction & $\approx$52{,}000 & Category, delta, balance, session-linked events \\
Session metadata & Session & $\approx$110 & Dates, durations, attendance/poll metadata \\
Perception survey & Response & 597 & 10 Likert + 2 open-ended items (Sec.~\ref{sec:methods}) \\
Later records (RQ4) & Student & 3{,}570 & Attendance, endorsements, projects, completion, 30 Jun--14 Sep \\
\bottomrule
\end{tabularx}
\end{table}

\subsection{Data Cleaning, Participation Variables, and Privacy}
A small number of learners appeared under two accounts, typically through primary and alternate email addresses. For the research dataset, duplicate persons were consolidated into the more complete record, removing 155 duplicate accounts from the student roster (7{,}945 to 7{,}790) and 110 from the started-learner population (3{,}717 to 3{,}607). The resulting participation segments were unchanged by this cleaning. The behavioural variables were designed to capture both the level and change of participation while accounting for differences in opportunities to participate. They comprised attendance rate, poll-participation rate, points earned per eligible session, four weekly point-velocity measures (weeks 1--4), and point acceleration from week 1 to week 2. Raw cumulative points were not used as the primary indicator because they are influenced by both participation and exposure time in a continuously enrolling programme. The behavioural dataset used pseudonymous identifiers throughout; the survey linkage is described in Section~3.3, and the survey itself was confidential rather than anonymous.

\subsection{Analysis}
The analysis was organised around the four research questions. For RQ1, participation trajectories were examined within the active learner population using the two-tier procedure described in Section~\ref{sec:design}. Learners were grouped using $K$-means clustering on the standardised trajectory and participation features, and the resulting segments were examined in terms of their weekly participation patterns and opportunity-normalised measures. Raw cumulative points were used descriptively where relevant but were not treated as the primary measure of participation.

For RQ2, learner perceptions were first summarised by segment using per-item descriptive statistics. The internal consistency of the Likert items was assessed using Cronbach's alpha~\cite{cronbach1951}, and an exploratory single-factor check was conducted for the composite. Differences across segments were examined using Kruskal--Wallis omnibus tests~\cite{kruskal1952} with Benjamini--Hochberg (BH) correction~\cite{benjamini1995}. Significant omnibus results were followed by Mann--Whitney $U$ post-hoc comparisons~\cite{mann1947} with Bonferroni correction and rank-biserial effect sizes. Epsilon-squared was used to estimate omnibus effect sizes~\cite{tomczak2014}. Spearman correlations~\cite{spearman1904} were also used to examine associations between perception items and their corresponding behavioural indicators, with BH correction applied to control for multiple comparisons.

For RQ3, the analysis focused on cases where recorded participation and learner perceptions did not align, particularly among active learners with low recorded participation. The two open-ended survey items were analysed with the two-cycle coding process described in Section~\ref{sec:design}. The coding process did not assume that low recorded participation reflected low motivation and was designed to identify both reported barriers and contradictory cases. For RQ4, segment membership fixed at the snapshot was cross-tabulated against each later outcome (chi-square tests with Cram\'er's $V$), and the area under the ROC curve (AUC) of the ordered segment label was compared with the AUC of cumulative points and of points per eligible session computed on the same snapshot. The RQ4 analysis was repeated for learners with a full four-week window before the snapshot (start on or before 1 June 2026) to check the effect of tenure. Throughout the analysis, attendance, poll participation, and points were treated as behavioural indicators of participation rather than direct measures of motivation, learning, or self-regulation.

\section{Results and Discussion}

\subsection{Results}

\subsubsection{Participation funnel and segments (RQ1).}
\begin{figure}[tbp]
\centering
\fbox{\includegraphics[width=0.46\textwidth]{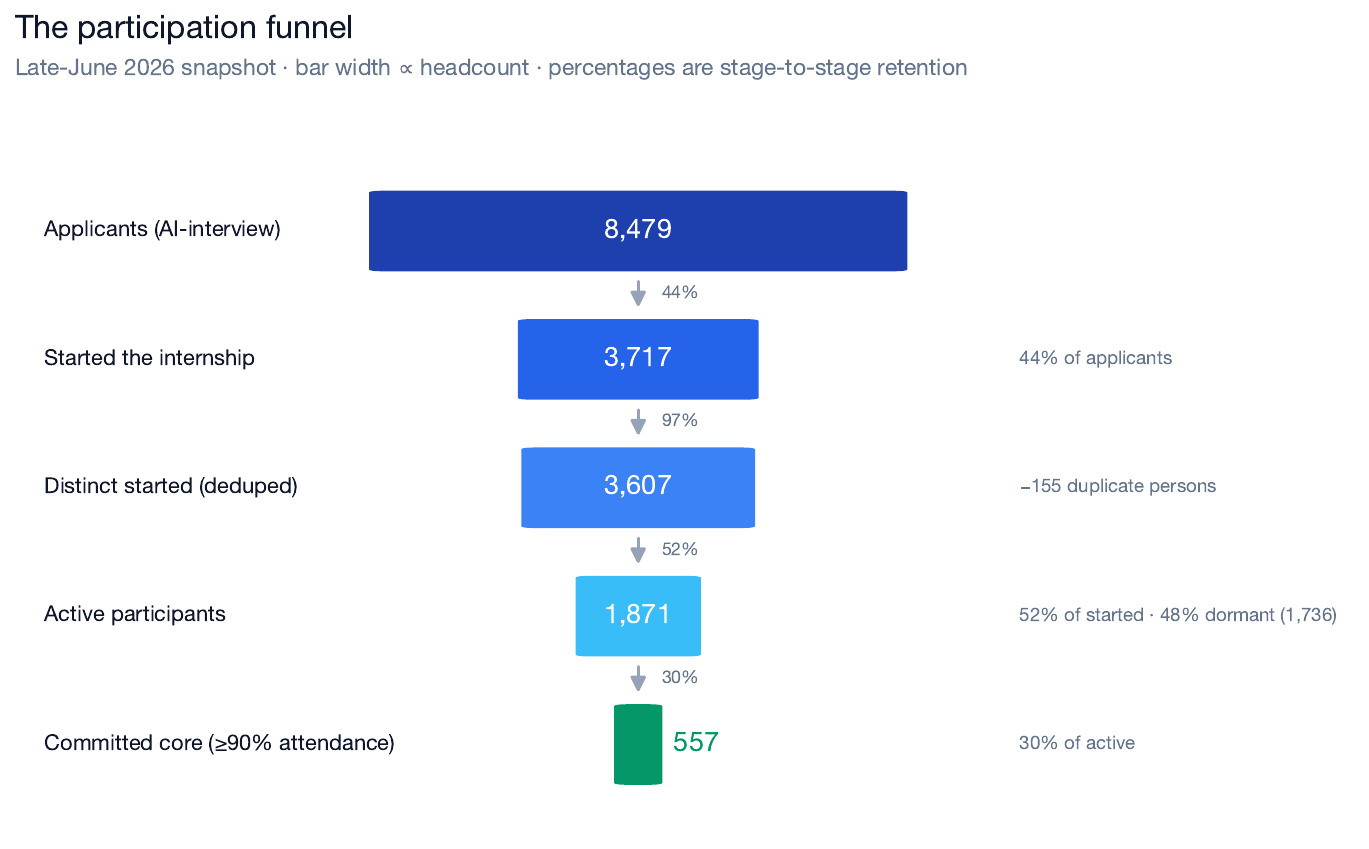}}
\caption{The participation funnel: 8{,}479 applicants $\to$ 3{,}717 started $\to$ 3{,}607 distinct $\to$ 1{,}871 active (52\%) $\to$ 557 committed core ($\geq$90\% attendance).}\label{fig:funnel}
\end{figure}

Of the 8{,}479 applicants who completed the admission interview by the snapshot, 3{,}717 had started the programme. After deduplication, 3{,}607 distinct learners who started remained. Tier~1 classified these learners into 1{,}871 active learners (52\%), who had at least one recorded attendance or poll response, and 1{,}736 dormant learners (48\%), who had started but had no recorded participation in either activity. Among the started learners, 557 attended at least 90\% of their eligible sessions during the observation window. The Thriving Core and Tapering segments together contained 784 learners, approximately 22\% of the started population (Fig.~\ref{fig:funnel}).

\begin{figure}[tbp]
\centering
\fbox{\includegraphics[width=0.5\textwidth]{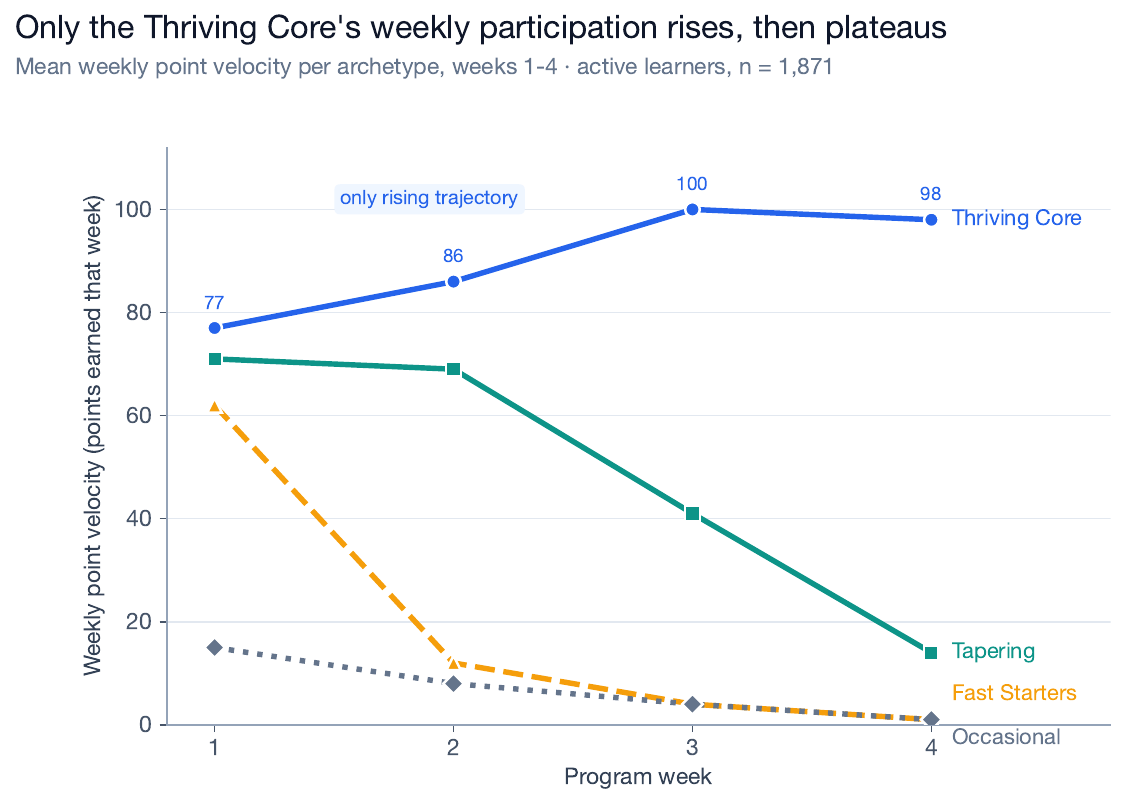}}
\caption{Mean weekly point velocity (points earned per week) for each participation segment across the first four programme weeks. Only the Thriving Core rises ($77\to98$); the Tapering, Fast Starter, and Occasional segments all fall.}\label{fig:velocity}
\end{figure}

Among the 1{,}871 active learners, $K$-means clustering of the normalised longitudinal participation features produced four segments ($k{=}4$, silhouette 0.44): Thriving Core ($n{=}497$), Tapering ($n{=}287$), Fast Starters ($n{=}303$), and Occasional Participants ($n{=}784$). The segments differed primarily in the shape and level of their participation trajectories (Fig.~\ref{fig:velocity}, Table~\ref{tab:fingerprint}). The Thriving Core showed increasing weekly point velocity across the first four programme weeks (77.1, 86.5, 99.6, and 98.5 points), an increase followed by a plateau. Tapering learners were strong in the first two weeks and then fell away (71 to 14 points per week), Fast Starters showed high initial participation followed by a sharp decline (62 to 1), and Occasional Participants showed low and intermittent participation. The Thriving Core had the highest attendance and poll-participation rates, with means of 0.96 and 0.90. Nearly half of the Tapering learners (47\%) had fewer than four observed weeks, compared with 7\% of the Thriving Core. These measures describe observed participation behaviour and should not be interpreted as direct measures of learner motivation or other psychological states.

\begin{table}[tbp]
\caption{Per-segment behavioural fingerprint (means on the normalised metrics used for clustering; separation is therefore expected by construction: the table's purpose is interpretability). The Dormant tier is rule-defined in Tier~1 and shown for reference.}\label{tab:fingerprint}
\centering\footnotesize
\begin{tabular}{@{}lccccc@{}}
\toprule
Metric & Thriving & Tapering & Fast Starters & Occasional & Dormant \\
\midrule
Attendance rate & 0.96 & 0.72 & 0.23 & 0.10 & 0.00 \\
Poll-participation rate & 0.90 & 0.63 & 0.17 & 0.05 & 0.00 \\
Points per eligible session & 16.4 & 11.3 & 3.1 & 1.1 & 0.0 \\
Weekly point velocity (wk 1$\to$4) & 77$\to$98 & 71$\to$14 & 62$\to$1 & 15$\to$1 & 0$\to$0 \\
$n$ & 497 & 287 & 303 & 784 & 1{,}736 \\
\bottomrule
\end{tabular}
\end{table}

\subsubsection{Perceptions, reasons, and barriers by segment (RQ2 and RQ3).}
The perception survey received 597 responses, of which 583 could be linked to the five started-learner groups. The six-item perception composite showed high internal consistency (Cronbach's $\alpha{=}0.914$); the six items loaded between 0.68 and 0.87 on a single factor, while the two negative-valence items formed a separate cost/barrier dimension. Across the full respondent group, top-two-box agreement was 87\% for staying on track, 83\% for motivation, 88\% for understanding the rules, and 84\% for regularly checking points. Fairness was weaker ($M{=}3.90$; 70\%), and points-related stress had the lowest mean ($M{=}3.24$), with 44\% selecting one of the two highest response categories. The ten-item Kruskal--Wallis analysis with BH correction found significant differences across segments for eight of the ten items, including helpfulness ($H(4){=}17.8$, $p_{BH}{=}.002$, $\varepsilon^2{=}.024$), understanding of the rules ($H(4){=}54.3$, $p_{BH}{<}.001$, $\varepsilon^2{=}.087$), motivation ($H(4){=}11.4$, $p_{BH}{=}.028$), and stress ($H(4){=}11.6$, $p_{BH}{=}.028$). Fairness ($p_{BH}{=}.39$) and barriers ($p_{BH}{=}.49$) did not differ across segments. The omnibus effects were small ($\varepsilon^2{\leq}.09$). Post-hoc differences were concentrated in comparisons involving the Thriving Core and Dormant learners (understanding of the rules, rank-biserial $r{=}0.50$). Spearman correlations between perception items and corresponding behavioural indicators were small when pooled ($\rho{=}.09$--$.14$) and close to zero within the two segments with adequate response coverage.

The open-ended barrier item produced 364 codable responses. Thirty-seven percent reported no barrier. Among responses reporting one or more barriers, the most common coded themes were infrastructural or technical barriers (38\%), including connectivity, power, and poll-submission failures; competing obligations (20\%); programme-design friction (14\%); and health-related barriers (3\%). No coded response explicitly attributed non-participation to disinterest. In responses describing how points affected participation, behavioural activation was the dominant theme (69\%), followed by self-monitoring or goal-striving (28\%); some responses also described affective or fairness-related costs (17\%). One learner described the goal-gradient effect of the leaderboard~\cite{kivetz2006} as ``a few more points and I may be there.'' Table~\ref{tab:joint} brings the two strands together by segment, with one learner's words for each segment and a convergence verdict.

\begin{table}[tbp]
\caption{Joint display: perceptions, reported barriers, and later outcomes by segment (behavioural profiles are in Table~\ref{tab:fingerprint}). Perception columns are item means (1--5; stress: higher $=$ more). Theme shares are of coded responses in that segment. Verdicts for the two groups with fewer than 20 respondents are exploratory.}\label{tab:joint}
\scriptsize
\setlength{\tabcolsep}{2pt}
\begin{tabular}{@{}>{\raggedright\arraybackslash}p{1.25cm}rcccc>{\raggedright\arraybackslash}p{2.3cm}>{\raggedright\arraybackslash}p{2.75cm}>{\raggedright\arraybackslash}p{2.0cm}@{}}
\toprule
Segment & $n$ & Helped & Motiv. & Fair & Stress & Themes (effect; barriers) & In their words & Verdict \\
\midrule
Thriving Core & 411 & 4.43 & 4.37 & 3.93 & 3.13 & activation 26\%, attendance push 22\%; no barrier 31\%, connectivity 28\% & ``the leaderboard ranking motivated me to not slip up in any polls'' & Converge: high participation, positive view, least stress; 94\% still attending \\[2pt]
Tapering & 101 & 4.22 & 4.19 & 3.89 & 3.47 & activation 29\%, attendance push 17\%; no barrier 25\%, connectivity 18\%, academic load 12\% & ``it does motivate me \ldots but I am becoming more focused on the points than on actual learning'' & Converge on level, diverge on direction: positive view, falling participation; 55\% still attending \\[2pt]
Fast Starters & 8 & 3.50 & 3.62 & 3.12 & 3.75 & attendance push, activation; academic load, technical & ``examinations and internet connectivity issues made it difficult to attend every session'' & Exploratory: lowest fairness, highest barriers; 7\% still attending \\[2pt]
Occasional & 18 & 3.78 & 4.11 & 3.89 & 3.22 & self-monitoring 33\%; no barrier 40\%, connectivity 20\% & ``system not recognising the id, my attendance was inaccurate in the starting'' & Exploratory: positive view, low participation; 4\% still attending \\[2pt]
Dormant & 45 & 4.07 & 4.02 & 3.84 & 3.60 & fairness concern 30\%, activation 21\%; other 26\%, connectivity 18\% & ``there could be `n' number of reasons for not attending \ldots need not mean the student is not interested'' & Diverge: motivating in principle, blocked in practice; 6\% attending later \\
\bottomrule
\end{tabular}
\end{table}

\subsubsection{Segments and later outcomes (RQ4).}
Segments were assigned using data available up to 29 June 2026. Outcomes were then obtained from platform records for the following eleven weeks, from 30 June to 14 September 2026, without changing the assigned segment labels. Table~\ref{tab:outcomes} reports, for each segment, the proportion of learners who attended at least ten additional sessions, received at least one additional endorsement on a peer-teaching question, submitted a project, completed a project, reached the Legend league, and completed the internship. Self-paced course completion in this period was rare (under 5\% in every segment except Tapering, 22\%) because most learners completed these courses during their first weeks in the programme; it is reported in the supplement.

\begin{table}[tbp]
\caption{Segments assigned by 29 June and outcomes recorded 30 June to 14 September 2026 (percent of segment; $n$ after the pseudonym join in brackets). All chi-square tests $p<.001$.}\label{tab:outcomes}
\centering\scriptsize
\setlength{\tabcolsep}{3pt}
\begin{tabular}{@{}lrcccccc@{}}
\toprule
Segment & $n$ & Attend.\ $\geq$10 & Endorsed & Proj.\ subm. & Proj.\ done & Legend & Completed \\
\midrule
Thriving Core & 497 (494) & 93.9 & 84.4 & 47.8 & 28.3 & 31.8 & 21.7 \\
Tapering & 287 (284) & 54.6 & 57.0 & 16.2 & 7.7 & 7.7 & 6.0 \\
Fast Starters & 303 (303) & 6.6 & 11.6 & 0.7 & 0.0 & 0.0 & 0.0 \\
Occasional & 784 (777) & 3.7 & 6.7 & 0.3 & 0.1 & 0.1 & 0.1 \\
Dormant & 1{,}736 (1{,}712) & 6.1 & 9.6 & 0.8 & 0.0 & 0.0 & 0.0 \\
\midrule
Cram\'er's $V$ & & 0.77 & 0.66 & 0.59 & 0.47 & 0.50 & 0.41 \\
\bottomrule
\end{tabular}
\end{table}

The segments differed substantially in subsequent behaviour across all six outcomes (all $\chi^2$ tests $p<.001$), with Cram\'er's $V$ ranging from 0.41 for internship completion to 0.77 for attendance. The Thriving Core showed the highest levels of continued participation: 93.9\% attended at least ten additional sessions and 84.4\% received at least one additional peer-teaching endorsement. The corresponding figures for Tapering were 54.6\% and 57.0\%, while fewer than 12\% of learners in any of the other three segments met either threshold. The same pattern was observed for later milestones. Project submission, project completion, reaching the Legend league, and internship completion were concentrated in the Thriving Core, with 47.8\%, 28.3\%, 31.8\%, and 21.7\% respectively. Tapering showed lower but non-zero rates for these outcomes, whereas the remaining three segments recorded almost no project completion, Legend attainment, or internship completion.

The separation in later behaviour was closely related to the amount of participation already observed by 29 June. Among active learners, the four-segment label predicted subsequent attendance with an AUC of 0.94, equal to the AUC obtained using cumulative points up to 29 June; points per eligible session produced an AUC of 0.97. For internship completion, the corresponding AUCs were 0.86, 0.86, and 0.89. Segment trajectory therefore provided limited additional separation beyond participation level overall. Its value was more apparent among learners with similar cumulative points. In the 200--300 point range earned over four weeks, 17\% of Thriving Core learners completed the internship compared with none of the other segments ($n{=}48$ and 34), suggesting that trajectory shape can distinguish learners whose overall participation level is similar.

The Tapering pattern also reflects differences in learner tenure. When the analysis was restricted to learners who had a full four-week observation window before 29 June (start date on or before 1 June), the proportion still attending in Tapering fell to 22\%, with no internship completions, while the other four segment outcomes remained unchanged. Learners who joined during the final two weeks of June had less time to demonstrate a sustained rise in participation and could therefore be classified as Tapering by construction. The segment should consequently be interpreted as indicating participation that was not yet sustained, rather than necessarily indicating a decline in engagement.

Two points constrain the interpretation of these outcomes. Internship completion requires 3{,}600 attended minutes and an accepted project, both of which also contribute to points. Internship completion is therefore not independent of the participation measure and should not be treated as an independent validation of the points system. Continued attendance and peer-teaching endorsements provide a closer test because they represent subsequent behaviour without being completion requirements. In addition, the outcome window covers only eleven weeks, so completions occurring after 14 September 2026 are not included.

\subsection{Discussion}

The results show that participation is not adequately described by a single cumulative point total. The four active-learner segments differed not only in the amount of participation but also in how participation changed across the first four programme weeks. The Thriving Core increased its weekly point velocity before reaching a plateau, whereas Fast Starters showed a sharp decline after an initially high level of participation. Tapering learners fell away more gradually, while Occasional Participants remained low and intermittent. This extends work on learner engagement trajectories~\cite{kizilcec2013,ferguson2015,coffrin2014} by showing how longitudinal patterns can also be observed in a gamified participation ledger, and it supports the view that time should be treated as a dimension of learner behaviour rather than reducing participation to a cumulative level~\cite{knight2017}. The outcome analysis adds a qualification. Over eleven weeks, the segment label and the cumulative points earned by the snapshot predicted later participation about equally well; the shape of the trajectory added information mainly where totals were ambiguous, in the band where learners with rising and falling trajectories held similar totals. A total is therefore a fair summary for learners at the extremes and a poor one in the middle.

The two-tier analysis also shows why participation status and participation pattern need to be considered separately. Nearly half of the learners who started had no recorded attendance or poll participation and were therefore classified as dormant before trajectory clustering. Including this group directly in clustering would mix the absence of observable participation with differences among learners who had observable trajectories. Separating the groups provides a more interpretable description of the data: learners who had not started, learners who had started without recorded participation, and active learners with different participation patterns. These categories describe observable programme behaviour and should not be treated as psychological types.

What the later records validate is persistence, not learning. Segments assigned at week four separated learners who kept attending, kept teaching peers, and reached later milestones from those who did not, with almost every later milestone concentrated in the Thriving Core. This is evidence that the segments are a valid early indicator of behavioural engagement in the sense of continued participation~\cite{fredricks2004}. It is not evidence about learning quality, because no outcome in the platform record measures learning directly, and the completion outcome is itself built from point-earning activity.

The perception results suggest that the points system was generally viewed positively, but that this experience varied across segments. Learners reported high usefulness, motivation, rule clarity, and self-monitoring, while fairness received a lower rating and points-related stress was reported by a substantial minority. The segment differences were statistically significant for several perception items and survived pooling of the small groups, although the omnibus effects were small. The low correlations between individual perception items and behavioural indicators are important: segment-level differences in perception should not be read as evidence that point-based behaviour predicts an individual learner's perceptions.

The joint display shows where the two strands agree and where they part. They converge for the Thriving Core and, on level, for Tapering. They diverge for the Dormant respondents, who described the points system as motivating while reporting external barriers, and the same pattern appears among the few responding Occasional and Fast Starter learners. Technical problems, connectivity, competing obligations, and programme-design issues were more prominent in the open-ended responses than explicit expressions of disinterest. Given the strong self-selection in the survey, this divergence is a hypothesis about external constraint among low-engagement learners, to be tested with a sampled survey, rather than a finding about the segments as a whole. It does, however, show that low recorded participation can coexist with positive perceptions of the incentive system~\cite{ryan2000}, so low recorded participation should not automatically be interpreted as low motivation.

Taken together, the findings support interpreting gamified participation data as behavioural indicators rather than direct measures of learner engagement. Their interpretation becomes more informative when participation is examined longitudinally and relative to learners' opportunities to participate, rather than through cumulative points alone. Behavioural traces and learner reports provide different types of information: behavioural data show what participation looked like and, from the outcome analysis, what it led to on the platform; learner reports provide possible context for why participation may have been limited. For programme monitoring, this supports using a declining or persistently low trajectory in the first weeks as a signal for further investigation rather than as a standalone judgement about learner motivation.

\subsection{Limitations}
The study is observational and makes no causal claims about gamification. Outcomes were observed for eleven weeks after segmentation. Completion is defined by point-earning activity and cannot validate the points on its own; continued attendance and peer endorsements are independent of the segmentation features in time but not in kind, and a learning outcome measured outside the platform was not available. The Tapering segment mixes learners who faded with learners who joined too late to show a rise; the full-window analysis bounds the size of that mixture. Identity-matching errors may affect some learners, although the segmentation was robust to duplicate-account cleaning. The clustering depends on feature and preprocessing choices; while stable across seeds and resamples, agreement with alternative algorithms was moderate, so the segments are not uniquely determined. Eighteen percent of active learners had fewer than four observed weeks, resulting in right-censoring of later velocity measures; the structure remained similar after excluding short-tenure learners (ARI 0.87). A re-registration of 741 learners on 30 June to 5 July reset their points and start date on the live platform; the 29 June export used here is unaffected, but analyses on the later export must recover first start dates. The survey is self-reported and strongly self-selected, with lower-participation segments substantially under-represented, limiting generalisation of perception results. Finally, the analysis focuses mainly on attendance and polls as structured behavioural traces, and the qualitative coding, although reliable on a 20\% double-coded subsample ($\kappa{=}0.88$--$0.91$), cannot capture all forms of learner engagement.

\section{Conclusion}
Cumulative point totals alone do not capture how learner participation changes over time. In this large-scale, continuously enrolling remote internship, normalised longitudinal trajectories revealed distinct participation patterns among active learners, while the two-tier model separated learners with no recorded participation before examining variation among those with observable participation. Segments assigned from the first four weeks predicted continued participation and later milestones over the following eleven weeks, with the shape of the trajectory adding to the cumulative total mainly where totals were ambiguous. Learner perceptions provided additional context, particularly where low recorded participation coexisted with reported motivation and external barriers. Together, these findings suggest that gamified participation data are more informative when interpreted as longitudinal, opportunity-aware behavioural indicators and considered alongside learner reports. Future work should test early trajectory patterns against learning outcomes measured outside the platform, examine under-represented low-participation learners with a sampled survey, and replicate the approach in other incentive-based learning programmes.

\begin{credits}
\subsubsection{\ackname} We thank the Vicharanashala Lab for Education Design, Indian Institute of Technology Ropar, for supporting this work.

The authors used a generative AI assistant to help restructure and edit drafts of the text and to format the manuscript. The authors designed the study, checked all analyses, results, and references, and take full responsibility for the content.
\subsubsection{\discintname} The authors have no competing interests to declare that are relevant to the content of this article.
\end{credits}

\end{document}